\documentclass[11pt]{report}

\usepackage[final]{pdfpages}
\usepackage{ucs}
\usepackage{subfigure}
\usepackage{color,graphicx}
\usepackage{wrapfig}
\usepackage{setspace}
\usepackage{version}
\usepackage{fullpage}
\usepackage{framed}
\usepackage[color]{changebar}
\cbcolor{blue}
\usepackage{program}
\usepackage[authoryear]{natbib}

\begin{document}

\begin{flushleft}

\Large

{\bf A mathematical model of tumor self-seeding reveals secondary metastatic deposits as drivers of primary tumor growth.}
\normalsize

Jacob G.\ Scott$^{1,2}$, David Basanta$^1$, Alexander R.A.\ Anderson$^1$ \& Philip Gerlee$^{3,4}$\\

$^1$ Integrated Mathematical Oncology, H. Lee Moffitt Cancer Center and Research Institute, Tampa, FL, USA\\
$^2$ Centre for Mathematical Biology, University of Oxford, UK\\
$^3$ Sahlgrenska Cancer Center, University of Gothenburg; Box 425, SE-41530 Gothenburg, Sweden \\
$^4$ Mathematical Sciences, University of Gothenburg and Chalmers University of Technology, SE-41296 Gothenburg, Sweden \\


Running title: Tumour self-seeding

\end{flushleft}


\section*{Abstract}
Two models of circulating tumor cell (CTC) dynamics have been proposed to explain the phenomenon of tumor 'self-seeding', whereby CTCs repopulate the primary tumor and accelerate growth: Primary Seeding, where cells from a primary tumor shed into the vasculature and return back to the primary themselves; and Secondary Seeding, where cells from the primary first metastasize in a secondary tissue and form microscopic secondary deposits, which then shed cells into the vasculature returning to the primary. These two models are difficult to distinguish experimentally, yet the differences between them is of great importance to both our understanding of the metastatic process and also for designing methods of intervention.  Therefore we developed a mathematical model to test the relative likelihood of these two phenomena in the subset of tumours whose shed CTCs first encounter the lung capillary bed, and show that Secondary Seeding is several orders of magnitude more likely than Primary seeding. We suggest how this difference could affect tumour evolution, progression and therapy, and propose several possible methods of experimental validation.

\newpage

\section*{Introduction}
Metastatic spread of cancer accounts for the lion's share of cancer associated death.  The transition from localized to metastatic disease also represents a therapeutic paradigm shift for patients and clinicians, as goals change from curative to palliative.  Regardless of the import of this change of state, we understand the mechanisms of this process very poorly -- beyond genetic correlations \citep{Bos:2009jl,Minn:2005ia} and some beautiful interrogation of specific points in the cascade \citep{Luzzi:1998dy,Chambers:2002co}, we are basically in the dark.  In the past several years our ability to measure and interrogate the vector of hematogenous spread -- the circulating tumor cell -- has begun to improve \citep{Marrinucci:2010go,Marrinucci:2007km,Stott:2010iw,Cristofanilli:2004hw,Cristofanilli:2005es}.  This improvement has yet to yield any appreciable clinical gains as we have not yet put these technologies into widespread clinical use; nor have we rigorously established or validated any theory in patients.   

One of the few exceptions to this is a series of papers beginning with the work by \cite{Norton:2006us} which first suggested the possibility of tumor self-seeding: the idea that tumor cells shed into the vasculature could end up coming back to the primary tumor to drive growth and progression.  This idea appealed to the imagination of theorists, scientists and clinicians alike and was then beautifully shown to exist in a paper by \cite{Kim:2009kv} -- in which mice were given two orthotopic breast tumors each tagged with a different fluorescent color, which were then shown to populate and promote growth in the contralateral tumor after two months.  This phenomenon was also shown in several other tumor types (including melanoma, colon and skin) and several other interesting biological insights were made that included gene expression analysis suggesting possible mechanisms for this preferential colonization.  

In the same issue of the experimental journal, Leung and Brugge provided an insightful review of the literature on the subject of self-seeding and the role of several tumor-derived cytokines (IL-6 and 8, among others) which were found to be upregulated in the ‘seeded tumors’ \citep{Leung:2009hb}.  They point out that these cytokines have been shown to be involved in everything from easing extravasation, to promoting tumor vasculature formation, and tumor relapse \citep{Bos:2009jl, Schafer:2007ev}. They also postulated that perhaps the self-seeding could occur not only from the primary directly back to itself but also through a route that included sub-clinical secondary metastatic deposits, which could then "communicate" with the primary tumor via their own shed progeny \citep{Leung:2009hb}.  This supposition, however, is very difficult to experimentally test, and is neither supported nor refuted by extant data in the literature.

This leaves two very distinct possible routes by which a circulating tumor cell (CTC) can promote the growth of the primary tumor. Both begin with the cells accessing the bloodstream, either by gaining the ability to actively intravasate through mutation or cytokine-driven transformation, or by exposure to flowing blood via endothelial disruption or tumor involvement in the vascular lining  \citep{Mazzone:2009kh, Chang:2000iv, Bockhorn:2007dx}, and then surviving as free floating CTCs.  The paths then diverge -- the first route, which we will call Primary seeding, then involves these CTCs avoiding arrest at intervening capillary beds, and  successfully extravasating back into the primary tumor.  

The second route, which we will call Secondary Seeding, begins in the same way, but involves several additional steps; after successful intravasation these cells arrest in a capillary bed that is not the primary tumor, which for most tumours would be the lung, extravasate and grow into a small colony.  Then, after some period of time, during which this secondary colony would have expanded from a single cell to a small colony ($\sim 10^6 -10^9$ cells), and been exposed to different evolutionary pressures, a cell or cells from this secondary colony would intravasate, circulate, and return to the primary tumor. In this case it is not the cells which originally left the primary that return, but their descendants having multiplied and evolved at secondary sites in the body. Nevertheless, the departing cells, through a chain of events, accelerate the local growth of the primary. It should be noted that these two routes need not be mutually exclusive, and they might simultaneously contribute to the growth of the primary tumor.

Teasing these two routes apart experimentally has been, to this point, extremely difficult, as the secondary deposit of cells need never become clinically meaningful, or for that matter, much more than a small colony, difficult even to detect on careful dissection, and certainly below the imaging threshold.  The onus therefore lies on theoreticians to attempt to understand the differences in likelihood between these two routes, and it is this burden that we attempt to shoulder in the remainder of this work. We will do this by constructing a model comprised of several different mathematical constructs, which captures the local growth of the primary, the dispersal of cancer cells into the circulation, and their return to the primary site.  We will also suggest several simple experiments that could be undertaken to help validate this model and several possible clinical interventions which could effect clinical practice.

\section*{Parameter estimation: A classic Fermi problem}
At the heart of any theoretical model are the parameters used to calculate the outcome of the model.  For most biologically inspired models, these parameters are drawn from the literature, usually averaged over many papers, or directly from experiments designed specifically for model creation.  In this paper, while there is some published research with measurements of our parameters, many of them remain cloaked in mystery, typically because we do not yet have the ability to measure these things reliably in human subjects.  This lack of data has not deterred us from attempting a reasonable parameterization of this model.  As physicists by training, we will proceed from this state of unknown by a tried and true method referred to as `Fermi estimation' -- named for the famed nuclear physicist Enrico Fermi, who was well known for this sort of formulation \citep{guess}.  Famously, at the Trinity nuclear test site, Professor Fermi dropped several torn up pieces of paper during the shock wave from the first ever nuclear explosion, and calculated, within an order of magnitude, the energy of the blast.  This type of formulation is based on a series of 'best guess' approximations that are in the end multiplied together.  On a log-scale the error of such an estimate typically scales with $\sqrt{n}$ where $n$ is the number of estimates in the chain. The reason for this being that the estimation chain can be viewed as a series of coin flips, where one either over or under estimates the given quantity, and such a process is described by the binomial distribution, whose standard deviation scales with $\sqrt{n}$, where $n$ is the number of coin flips or guesses. This implies that if we make an approximation in each step which is within one order of magnitude, i.e.\ if the true value is $1$ then our guess is in the range $0.1-10$, then the error that we make will be $E = 10^{\sum_i^n x_i}$, where each $x_i=1$ or $-1$. Now the sum is distributed according to a binomial distribution whose standard deviation scales as $\sqrt{n}$, meaning that $\log E$ will typically scale as $\sqrt{n}$. For example, for a Fermi estimate made with 9 steps, each within one order of magnitude, the error will be approximately $E \sim 10^{\sqrt{9}}=10^3$.

\begin{figure}[ht]\label{fig:schematic}
\centering
\subfigure[Primary Seeding schematic]{
\includegraphics[scale=.4]{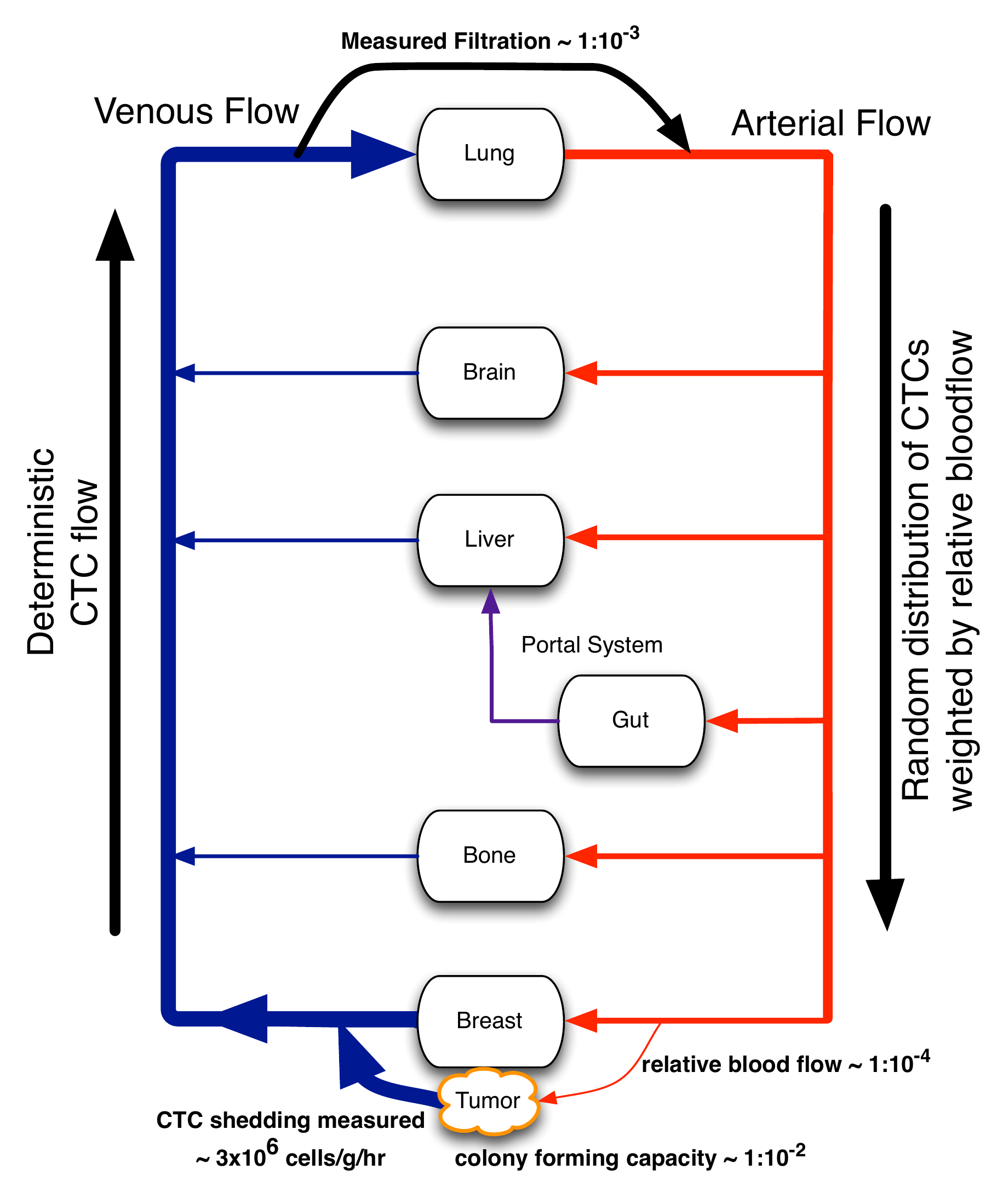} \label{fig:filterflowa}}
\subfigure[Secondary Seeding schematic]{
\includegraphics[scale=.4]{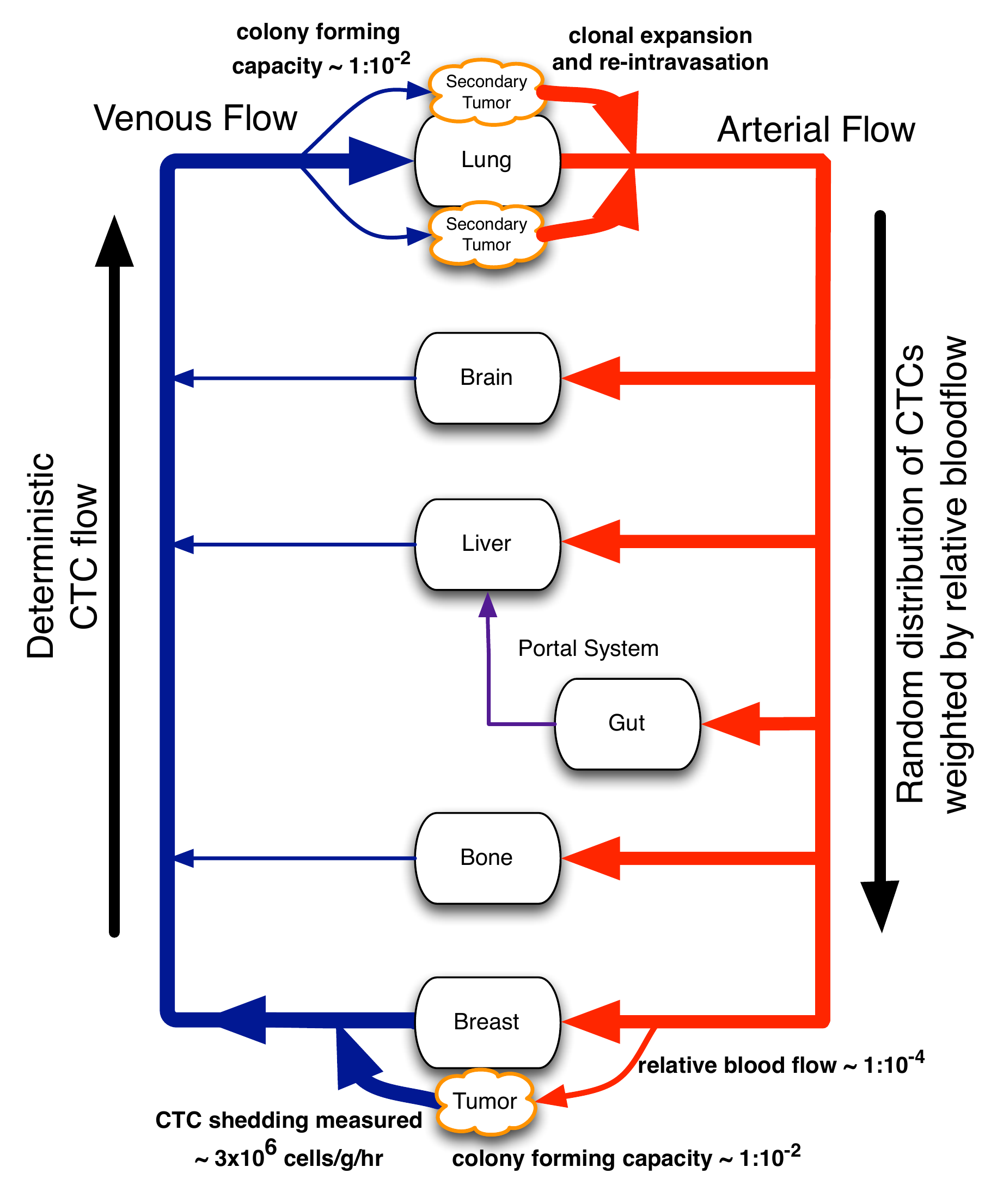} \label{fig:filterflowb}}
\label{fig:filterflowschematics}\caption{A schematic representation of a patient with a tumor in the breast showing anatomic structures as nodes in a network with physiologic vascular connections as edges. Thickness of vascular connections correlates with relative number of CTCs on a log scale.}
\end{figure}

With this in mind, let us now try to approximate the probability of a single cancer cell returning to its site of origin by estimating the likelihood of the most important steps necessary for this to occur, and from these form a Fermi estimate of the probability of the entire process. We will  consider the probability of each route in turn, beginning with Primary Seeding, illustrated schematically in figure\ \ref{fig:filterflowa}.

\subsection*{Primary Seeding}
For the sake of simplification, we choose a primary breast cancer for our example, and begin the physical process of Primary Seeding in the post-capillary venule, where a cell shed from a tumor might enter the bloodstream.  We assume that all cells in the primary tumor are able to access the bloodstream and remain viable, which is likely a vast overestimate in the favor of Primary Seeding \citep{Bockhorn:2007dx}. This journey is undertaken by millions of cells (up to $4\times10^6$ cells shed/gram of tumor/day) \citep{Butler:1975ws}, each of which is subjected to slow, deterministic, one way flow in the venous blood, until it reaches the capillary bed of the next organ in its path -- the liver for tumors of the mid and hind-gut and the lung for all others (with the exception of some primary lung cancers).  It is for CTCs that reach the lung as the first capillary bed that we will focus our analysis in the remainder of this work.  A similar analytical treatment could be applied to other tumours, but, as CTCs shed from their micrometastases would be reduced by filtration during passage through the lung, our results can not be generalized.


While not rigorously quantified, the fate of tumor cells arresting at end organs has been elucidated by various groups with different specific measurements, but with a general consensus of an approximate 3 order of magnitude reduction \citep{Okumura:2009hr}, meaning that about 1 in 1000 cells passes through. If a cell originating from a non-hindgut or lung malignancy does manage to pass through the first `filter', which in the case we consider is the lung capillary bed, the environment changes from the slow, low pressure and low shear flow of the venous blood to the high pressure and high shear flow in the arterial system.  At this stage, the remaining cells in circulation are swept stochastically with the blood into arterial tributaries, most likely equally in proportion to the fraction of blood that goes down each one.  At this point, to get ‘back home’ via the bloodstream, a cell has at best a chance equal to that of the relative blood flow to the tumor as compared to cardiac output -- approximately the ratio of tumor mass to whole body mass ($\sim$10g/100kg = $10^{-4}$). For simplicity we will assume that all cells which pass through the vasculature of the primary successfully extravasate. If a cell was not 'lucky' enough to go down the right path leading to the tumor, it would encounter the capillary bed of another foreign organ with its filtration characteristics, followed by another pass of the entire system, starting again with the lung capillary bed.  Because of the strong reduction in cell numbers associated with this, we will consider a cell's chances in its first pass as a reasonable approximation of the probability of returning to the primary site.

It has been argued that primary tumors might emit chemotactic signals, such as interleukins \citep{Norton:2008uy}, which inform the circulating cancer cells about the location of the primary, increasing the rate of extravasation, and hence giving rise to a higher return probability. Since we have assumed that all cells which pass by the primary tumor extravasate, any such signal would need to alter the fate of the cells as they are traversing the arterial side of the circulatory system. This does, however, seem highly unlikely, since the transport of these cytokines is driven purely by diffusion, with diffusion constants on the order of $10^{-9} \mathrm{\ cm}^2/\mathrm{s}$ \citep{Poplawski2007}, while the velocity of the blood flow in capillaries is on the order of 0.1 cm/s \citep{Stuecker1996}. The dynamics of the cytokines are hence dominated by convective forces which results in limited upstream transport, making it nearly impossible for cells far upstream to sense the location of the primary, much less to 'home' toward it.

The final result of this back of the envelope calculation is that -- in the best case scenario -- 1 in 1000 cells ($10^{-3}$) make it through the first ‘foreign’ capillary bed and on the order again of 1 in 10,000 cells ($10^{-4}$) probabilistically end up ‘home’.  Simply getting back to the primary tumor does not guarantee success once there, in fact there is mounting data that supports that only specific side populations (often called cancer stem cells) are capable of forming viable colonies, and that these cells comprise approximately 1 in 100 ($10^{-2}$) cells out of the tumor bulk \citep{AlHajj:2003vn}.  All this taken into account,  at best around 1 in $10^9$ cells ($10^{-3} \times 10^{-4} \times 10^{-2} = 10^{-9}$) which successfully enter the bloodstream have the chance of contributing to future growth of the primary tumor growth via Primary Seeding.

\subsection*{Secondary Seeding}
For the second route, which we will call Secondary Seeding, the route (and associated calculations) remain essentially the same except that the circulating cancer cells skip two levels of 'filtration' in end organs, see figure \ref{fig:filterflowb}.  The tumour cells intravasate at the primary site and travel in the circulation, but once the cells reach the first foreign capillary bed, which for most tumours would be in the lung, we instead consider the fate of those that arrest and then subsequently survive extravasation.  At this point, however, the limiting step is survival and colony formation at the secondary site, before the beginning of the remainder of the journey. If colony formation is successful (estimated to 1 in $10^2$ by \cite{Luzzi:1998dy}) then a single cell can form a micro-metastatic colony consisting of $\sim 10^9$ clonal cells, which would likely escape clinical detection - indeed, many patients have been shown to have vast numbers of CTCs with only 'premetastatic lesions' suggesting the possibility of yet undocumented secondary colonies already in existence \citep{Husemann:2008iv}. The cells forming this putative secondary tumor all have the capability of extra- and intravasation from pre-existing mutations carried by the ancestral CTC.  Once these daughter cells begin to intravasate, at rates on the order of $\sim 10^6$ cells per hour (as per our primary tumor assumption), they are now subject to the same dispersal dynamics as the cells in the Primary Seeding example after their first 'filtration' step, that is about 1 in 10,000 have a chance to return to the primary tumor, and have a 1 in 100 possibility of contributing to primary tumor growth \citep{AlHajj:2003vn}.  The main differences here are that the cells participating in Secondary Seeding essentially ‘skip’ the filtration step, with its $10^3$-fold reduction, by extravasating at the lung capillary bed.  After this step, the estimated gains from clonal expansion are roughly speaking balanced by the unlikelihood of successful extravasation making for no change in return probability.  A possible benefit to this step, however, is that these cells could acquire mutations in a whole new fitness landscape, essentially widening their heterogeneity -- before beginning the trip back to the primary. 

\subsection*{Error estimation}
Before we proceed to describing a mathematical model of self-seeding we will try to calculate the errors made in our Fermi estimates. In Primary seeding we estimated the number of rate limiting steps to be three: 
 filtration, relative blood flow and colony formation. If we assume that each step is estimated within one order of magnitude, then according to the previous argument, the typical error of our final estimate will be $10^{\sqrt{3}}$ or $\sqrt{3} \approx 1.73$ on a logarithmic scale.  In the case of Secondary seeding we have four steps in the estimation chain: arrest, expansion, relative blood flow and colony formation.  By the above method then, the typical error of our final estimate will be $10^{\sqrt{4}}$ or $\sqrt{4} = 2$ on a logarithmic scale. 

\section*{Methods}
\subsection*{Mathematical model of self-seeding}
In order to investigate the importance of self-seeding in driving tumour progression, we have devised a model which captures the two main features of this process: local growth and dispersal. In an effort to simplify the system we make the assumption that the returning cells gain a growth advantage by rejoining the primary at favourable sites where cell division is enhanced. In fact this must be the case, at least to some degree, if self-seeding is going to accelerate tumour growth. If not, the cells would be better off staying in the primary and the net effect of cells leaving the primary would be a decrease in tumour mass. 
In line with \cite{Norton:2006us}, we thus assume that the primary tumor consists of a number of independent loci formed by returning cells, which together constitute the primary tumor. In each locus growth is assumed to follow logistic growth, which is similar, and in most cases indistinguishable  \citep{Winsor1932} from the Gompertzian growth law used by \cite{Norton:2006us}. It is however crucial that each locus is limited by some maximal size, otherwise self-seeding could never contribute to the growth of the primary.

\begin{figure}[!htb]
\begin{center}
\includegraphics[width=12cm]{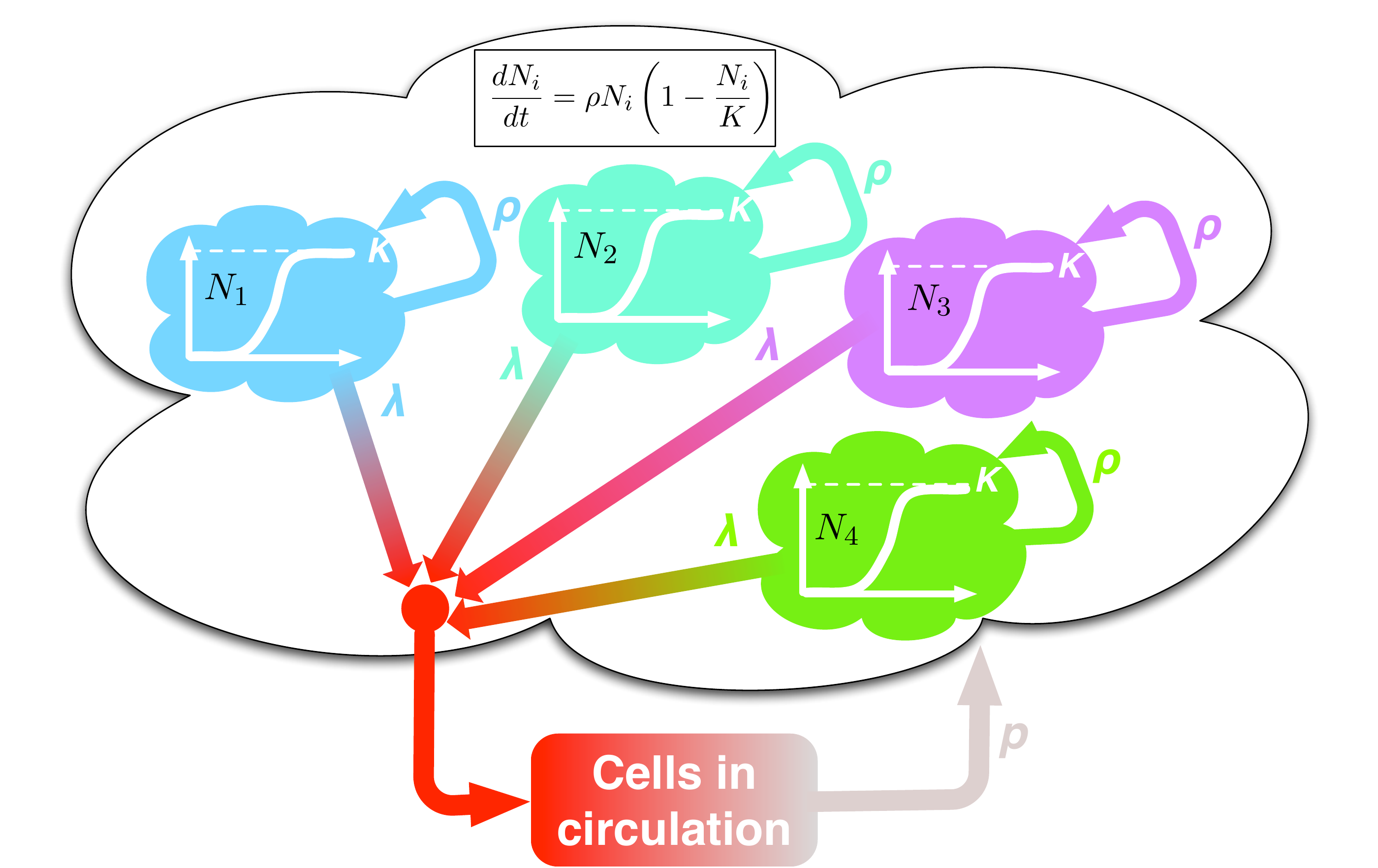}
\caption{\label{fig:fig2}{A schematic of the model of primary seeding. The primary tumour is assumed to consist of a number of independent loci, which grow according to the logistic equation (inset) bounded by a local carrying capacity $K$. Each of the loci release cells into the circulation at rate $\lambda$, and out of these circulating cells a fraction $p$ return to form new loci at the primary site. Starting with a single cell forming one locus the dynamics of the model generate an increasing number of loci, but the precise behaviour depends on the parameters of the model (see fig.\ \ref{fig:fig3}).}
}
\end{center}
\end{figure}


If we now let $n(t)$ denote the (integer) number of sites or loci at time $t$ that constitute the primary, then the dynamics of the model is described by $n(t)$ equations of the form

\begin{equation}\label{eq:log}
\frac{dN_i}{dt} = \rho N_i(1-N_i/K)
\end{equation} 
where $N_i$ is the size of the $i$th locus, $\rho$ is the growth rate and $K$ is the local carrying capacity. For simplicity we assume that the two parameters $\rho$ and $K$ are equal for all loci. The dynamics of each locus is initially exponential (when $N_i/K \ll 1$), but due to competition for resources it is bounded, and reaches a steady-state when the population size reaches the local carrying capacity ($N_i = K$). Solving these equations gives us information about the size of each locus $N_i(t)$ at time $t$, but we also need to prescribe the dynamics of the seeding process. In order to account for this process, which involves single cells, we proceed as follows: at each time step of the model, which corresponds to 24 hrs, we calculate the number of cancer cells from each locus that leave the primary. We assume that cancer cells leave the primary at a rate $\lambda$ (day$^{-1}$), and that the leaving events are independent of each other. This means that we can model the number of cells leaving locus $i$ by drawing a random number $r_i$ from a Bernoulli distribution with parameters $N_i(t)$ and $\lambda$. The number $r_i$ corresponds to the number of positive outcomes in $N_i(t)$ independent random trials, each successful with probability $\lambda$, i.e.\ we assume that all cells have an equal chance of entering the blood stream, an obvious over-estimate.

The number of cells remaining at each locus is adjusted according to $N_i(t+1) = N_i(t)-r_i$, accounting for the cells that entered the blood stream. The total number of cells leaving the primary at a given time $t$ is simply given by the sum of the contributing loci, i.e.

\begin{equation}
R(t) = \sum_{i=1}^{n(t)} r_i.
\end{equation} 

Secondly, we need to calculate the number of cells that manage to make it back via the circulation to the primary. Again we assume that these events are independent and that each circulating cell makes it back with a probability $p$, so that the total number of returning cells is given by a random number $m$ drawn from a Bernoulli distribution with parameters $R(t)$ and $p$. This means that we assume that the cells return instantaneously, which is a reasonable approximation since the growth dynamics occur on a much longer time scale than the circulation of cells.

The returning cells are all assumed to form new loci, meaning that at time $t+1$ we have $n(t+1)=n(t)+m$ equations of type (\ref{eq:log}) to consider. The initial condition for these newly formed loci is taken to be $N_i(t+1)=1$, for $i=n(t)+1,...,n(t) + m$. To simulate the model we thus alternate between numerically solving the logistic equation for each locus, and stochastically removing cells from existing loci and forming new ones. The total tumour burden is given by the sum over all loci and equals

\begin{equation}
N(t) = \sum_{i=1}^{n(t)} N_i(t).
\end{equation} 
The model contains four parameters: $\rho$, $K$, $\lambda$ and $p$. The growth rate is set to $\rho=\log 2$, giving an initial doubling time of each locus equal to 24 hrs, and the carrying capacity is set to $K=10^8$ cells, corresponding roughly to a spherical locus of diameter 1 cm. A reasonable estimate of the leaving rate $\lambda$ can be arrived at by considering the number of cells shed from a tumour. This number has been experimentally determined to be approximately $10^6$ cells gram$^{-1}$ day$^{-1}$, and since each gram of tumour tissue roughly contains $10^9$ cells, we arrive at an estimate of $\lambda = 10^{-3}$ day$^{-1}$. Again we invoke the error estimate discussed above, and with two steps and an error in each step of one order of magnitude, we end up with a typical error of $10^{\sqrt{2}} \approx 25$. The final parameter $p$, is the one with the largest degree of uncertainty, but as we have argued above it is most likely on the order of $10^{-9 \pm \sqrt{3}}$  for primary seeding, and $10^{-6 \pm 2}$ for secondary seeding.

\section*{Results}


To understand the role of tumor self-seeding, we used  a model similar to the one proposed by  \cite{Norton:2006us}, in which the primary tumour is assumed to consist of a number of independent loci (formed by returning cells), where the growth of each tumor locus is governed by logistic growth, which initially is exponential but is bounded by a local carrying capacity $K$ representing competition and limited nutrients.  We then incorporated this model into a larger model in which the cells in each locus shed stochastically into the blood stream at a rate $\lambda$ (day$^{-1}$) and return to form a new locus at random, each with probability $p$. The model is initiated with a single cancer cell forming one locus, and as time proceeds new loci are formed, each one growing logistically, giving a total tumor mass possibly larger than one achieved by non-seeded growth (see figure \ref{fig:fig2} for a schematic). The difference between primary and secondary seeding in the model is realised through a change in the parameter $p$, controlling the return probability of the cells. In secondary seeding the circulating tumor cells skip a filtration step, and also have the possibility to expand their clone at secondary sites in the lung, which in the model corresponds to an $10^3$-fold increase in the return probability \citep{Okumura:2009hr}.

Inspection of the vascular network (cf. fig.\ \ref{fig:filterflowa}) reveals that there are three common primary tumor types when it comes to CTC shedding dynamics: those originating in the lung which give rise to CTCs which can be immediately shed into the arterial vasculature; those originating in the gut which give rise to CTCs which first encounter the liver; and all others, which give rise to CTCs which first encounter the lung capillary bed.  While our analysis could be applied to any of these cases, the results would differ in each case.  
Here, we consider the latter case, and use the above model in order to investigate under which conditions (parameter values) primary and secondary self-seeding can give rise to accelerated tumour progression.
Note that in formulating the model we have made two assumptions in favor of the Primary seeding mechanism. Firstly, we have assumed that all cells in the primary tumor have an equal chance of entering the blood stream, while in reality only cancer cells adjacent to blood vessels have this capability, and secondly, that all returning cells end up in an new location adjacent to the primary, where growth can occur exponentially. In reality, this might be true for some returning cells, but a non-zero fraction will end up in locations already occupied by cancer cells, and these locations do not allow exponential expansion of the returning cells lineage. Taken together, this means that if the model does not show accelerated growth due to self-seeding under a certain set of conditions, then these are very unlikely to support accelerated progression {\it in vivo}.

\begin{figure}[!htb]
\begin{center}
\includegraphics[width=14cm]{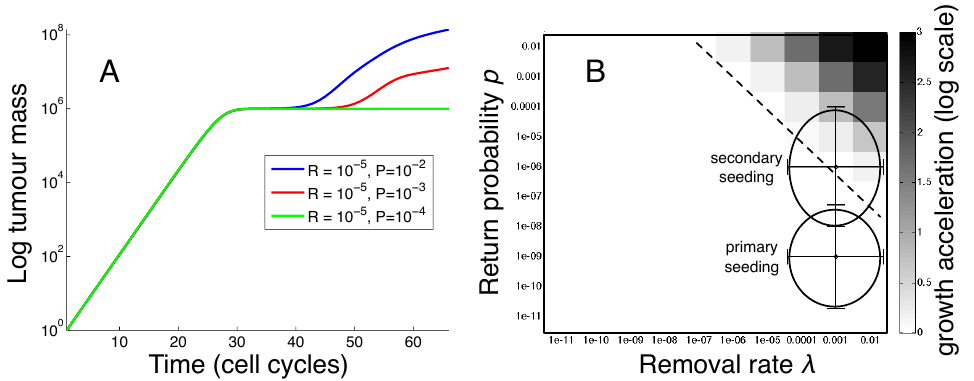}
\caption{\label{fig:fig3}Simulating the dynamics of primary seeding. (A) shows the total tumour burden for three different conditions where the removal rate was fixed at $\lambda=10^{-5}$ and return probability was taken to be $p=10^{-2},10^{-3}$ and $10^{-4}$ respectively. (B) illustrates the model dynamics when the parameters $\lambda$ and $p$ are varied systematically, and shows that accelerated tumour growth only occurs for large values of the leaving rate $\lambda$ and the return probability $p$. The parameter regions corresponding to primary and secondary seeding are encircled in the figure, and the error bars show the estimated error in the Fermi estimate. These results suggest that only secondary seeding has the capability of accelerating tumor progression, while primary seeding occurs with rates that do not alter the rate of tumour progression.} 
\end{center}
\end{figure}

Three different scenarios are shown in figure \ref{fig:fig3}A, with the removal rate fixed at $\lambda=10^{-5}$ and the return probability equal to $p=10^{-2},10^{-3}$ and $10^{-4}$. During the initial phase ($t < 40$ days) the starting locus grows exponentially (a straight line in the semilog-plot), but then settles down at the carrying capacity $K=10^6$, where a lower value of $K$ was used for the purpose of illustration. Only when the return probability $p$ is sufficiently high does the seeding mechanism give rise to new loci, which leads to significantly increased tumour growth. 

In order to investigate the range in parameter space in which the seeding mechanisms could affect tumor growth, we performed a systematic parameter exploration in which the two parameters related to seeding, $\lambda$ and $p$, were varied over nine orders of magnitude ($10^{-11} - 10^{-2}$). The two routes to self-seeding differ primarily in their likelihood of returning cells to the primary site, and hence in the parameter $p$ in the model. As a metric of tumor progression we measured the total tumour burden after 50 days from tumour initiation. Since the model is stochastic and contains an element of chance the results were averaged over 50 simulations for every parameter configuration. The result of these simulations is displayed in fig.\ \ref{fig:fig3}B, and shows that self-seeding can only lead to accelerated tumour growth in a small, isolated corner of parameter space. 

The conditions corresponding to primary and secondary seeding are denoted in the figure, and clearly the route of secondary seeding lies much closer and partly overlaps with the region in which accelerated progression is possible (upper right corner).
In fact, for a realistic removal rate of $\lambda = 10^{-3}$ day$^{-1}$ ($10^6$ cells shed/gram/day and each gram containing  $\sim 10^9$ cells) the smallest return rate which results in accelerated growth (defined as a total tumour burden which is larger than the local carrying capacity) is approximately $10^{-5}$, close to the estimate we arrived at for secondary seeding.

Taken together, these results suggest that although both processes might occur simultaneously, Primary seeding is highly unlikely to be an active contributor to tumour progression, while Secondary seeding seems to be a more probable candidate. The reason for this is that the secondary route allows for (but does not guarantee) a considerably higher return probability. 
As one can imagine, this step is subject to a large degree of stochasticity, which allows for significant inter-patient heterogeneity, and in some instances, this mechanisms is likely to provide tumor growth advantage, but certainly not all.  

\section*{Discussion}

\begin{figure}[ht]\label{fig:cartoon}
\centering
\subfigure[Primary Seeding cartoon]{
\includegraphics[scale=.4]{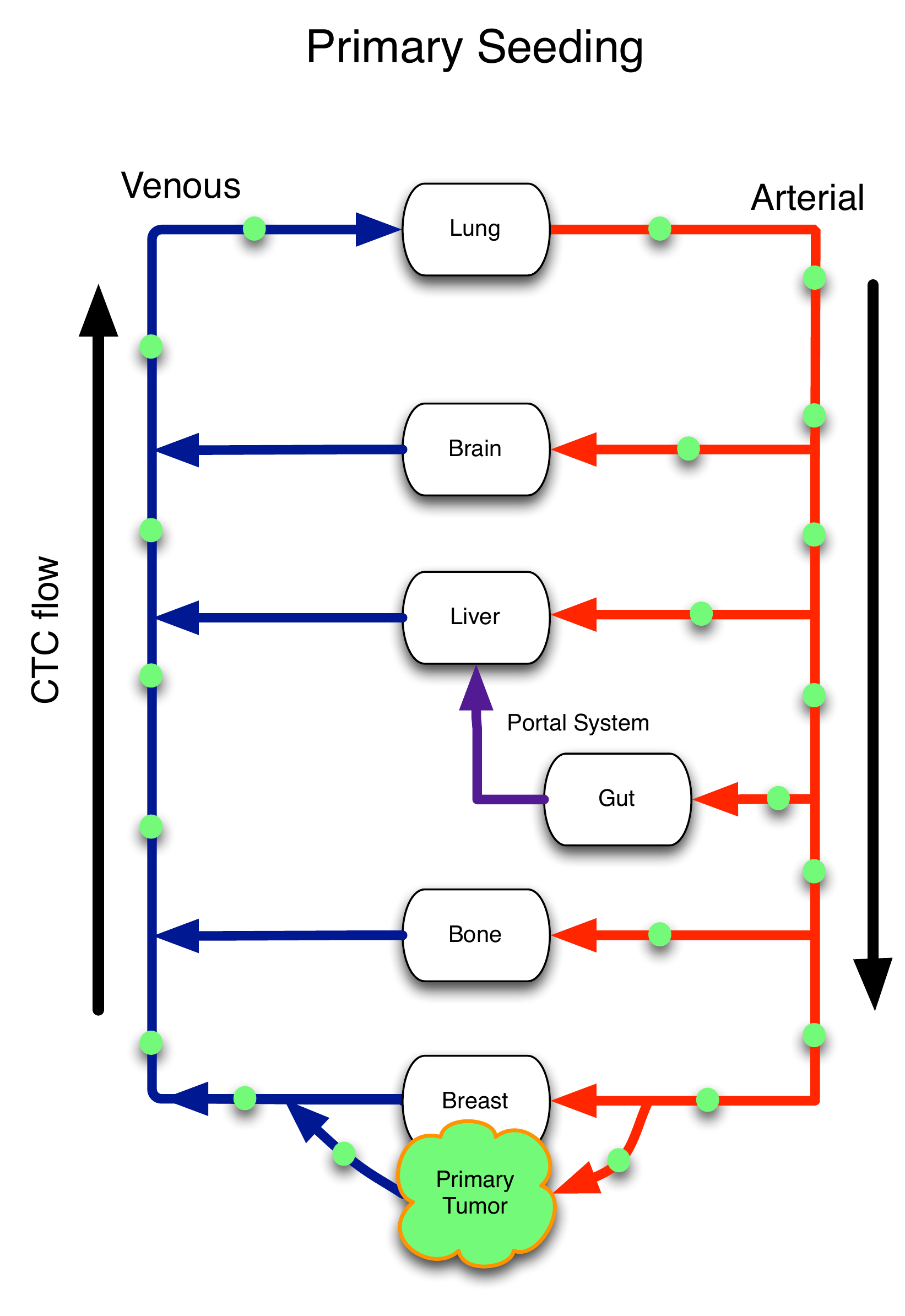} \label{fig:primarycartoon}}
\subfigure[Secondary Seeding cartoon]{
\includegraphics[scale=.4]{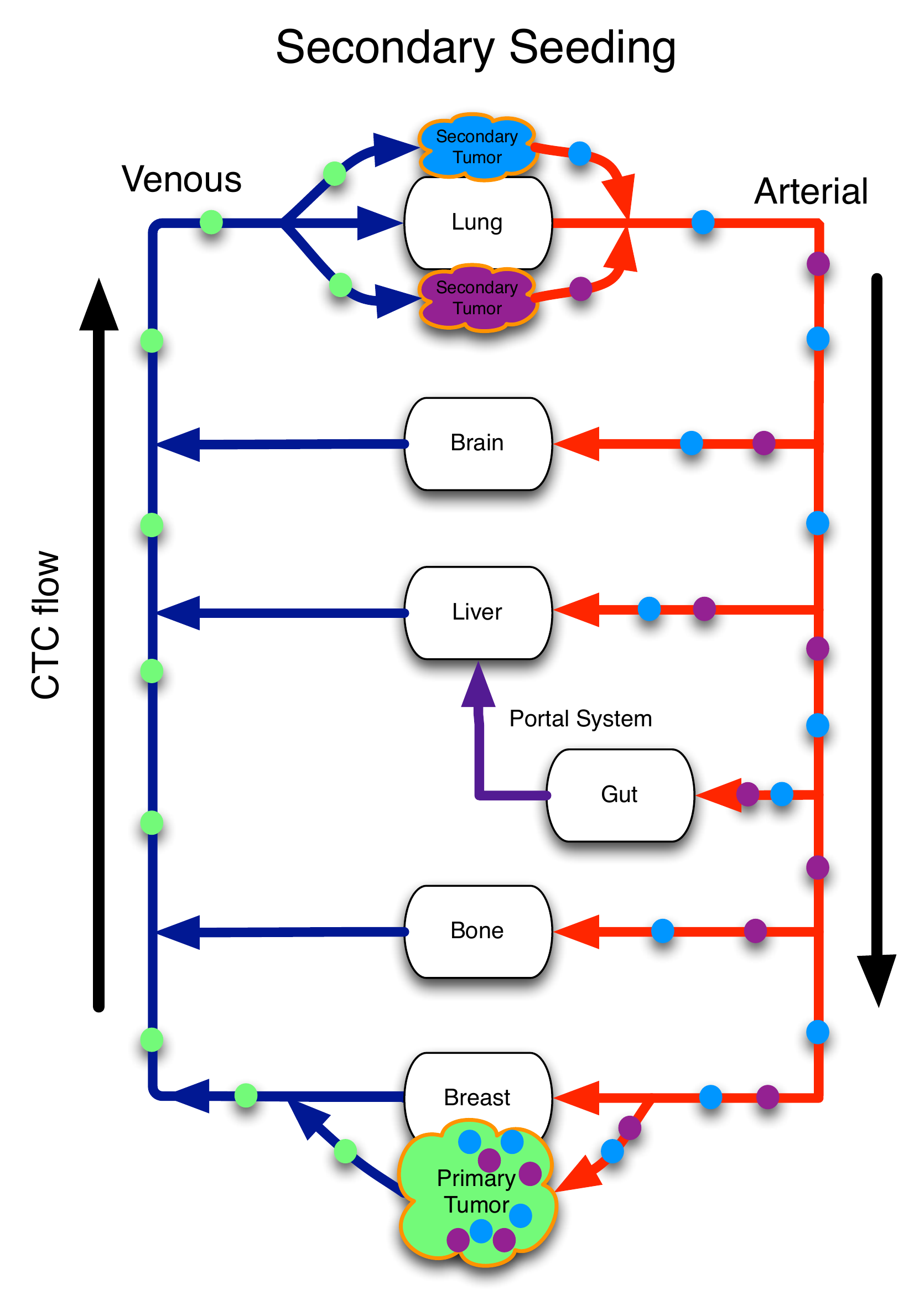} \label{fig:secondarycartoon}}
\label{fig:comparisoncartoon}\caption{A schematic representation of the different possible routes of self-seeding for (a) primary seeding and (b) secondary seeding.  On the left, we see the originally suggested route, which we call Primary Seeding, in which cells from the primary tumour (green) circulate through the vasculature and return to the primary tumour, which is the only tumor present in the body.  On the right, we see the second possible route, which we call Secondary Seeding, in which cells from micrometastases (purple and blue) are the ones to return to the primary. Note that for most tumors secondary seeding would involve lung micrometastases, but tumours arising from the node labelled 'gut' in this schematic represent a special case not included in our analysis.}
\end{figure}

Until only recently, measurement and characterization of the 'fluid phase' of metastasis has been experimentally intractable, and even now, the science remains in its infancy. The model of tumor self-seeding presented in this paper therefore explores experimentally untested ground with an admittedly imperfectly matched parameter set; despite this, it can provide answers to questions relating to the role of CTCs in driving primary progression, and more importantly focus future biological investigations in this regime. Experimental progress is also being made by several groups utilizing different methods \citep{Marrinucci:2007km, Marrinucci:2010go, Stott:2010iw}, which gives hope to further our understanding of this enigmatic phase of cancer progression in which the vector of hematogenous metastasis (the CTC) disseminates and gives rise to incurable disease. This characterization then, is of great import, not only because of the implications to our understanding of the process, but also because of the implications it presents to treatment and prevention of metastasis.  Further, many of the parameters which we have been forced to estimate, should be measurable once techniques to measure CTCs in different parts of the vascular system have been perfected and made safe.  Measurements such as these could also distinguish, in a patient by patient manner, if either Primary or Secondary Seeding was possible based on arterial vs. venous CTC concentrations (cf. figure \ref{fig:cartoon}).

These results become even more germane when the clinical reality of metastasis is considered.  When metastatic disease is confirmed in the clinic, removal of a primary tumor is typically relegated to palliative status.  There are certainly situations where resection of the primary can extend life -- if the tumor is invading a critical structure (great vessel, trachea) or causing GI obstruction, however in these cases the resection is not typically an 'oncologic resection' (i.e. en bloc, with wide margins of healthy tissue surrounding the tumor), but instead a minimal one designed to ease the symptoms and prevent impending disaster.  This paradigm has begun to be challenged however: first by a clinical trial showing a survival advantage in patients with metastatic renal cell carcinoma who received nephrectomy \citep{Flanigan:2001fk}; and now in a currently enrolling clinical trial for women with metastatic breast cancer who are being randomized to resection of the primary plus standard of care vs. standard of care alone (Eastern Cooperative Oncology Group trial E2108).  The putative mechanism for this benefit has been ascribed to immune benefits and the possibility of the primary tumor providing more efficient future metastasis as compared to extant metastatic deposits \citep{Danna:2004uq}, the latter of which our results lend a theoretical grounding to. 

Since the original statement of the Primary seeding hypothesis, there has been an explosion in our understanding of the genetic makeup of tumors at the single cell scale, allowing for better understanding of the spatial organization of genetically different clones.  Indeed, several recent papers have shown that individual tumors are made up of genetically different clones that have: evolved from the same precursor \citep{Gerlinger:2012fs}; and that are individually responsible for metastases \citep{Navin:2011gp}.  The reasons for this intra-tumoral heterogeneity are yet to be unravelled, although it has been shown computationally to be a consequence of differential microenvironmental selection pressures \citep{Anderson:2006wo} and is likely correlated with tumor grade -- with higher grade (more dedifferentiated) tumors being more heterogeneous.

To explain this heterogeneity, previous authors have used Darwinian evolution as a guiding principle which has been applied in numerous ways.  Additionally, authors have turned to the cancer stem cell hypothesis \citep{Sottoriva:2010ua} to explain how a somewhat smaller number of mutations could cause the same eventual heterogeneity in the tumor bulk. We offer a third suggestion which takes both previous ones into account and can help to further demystify the speed with which tumors develop resistance to targeted therapies.  We suggest that the far more probable Secondary Seeding route, in which lung metastases feed the primary tumour, can account for a faster route towards a heterogeneous primary tumor.  To augment the speed of a tumor's widening heterogeneity, this route offers two advantages:  it contains a mechanism by which a tumor can explore many different fitness landscapes (different organs), and also a mechanism by which cells are subject to differential stresses and physical selection processes in the vasculature \citep{scottfilterflow}.  The consequences of the former, a widened evolutionary space in which to search, is the subject of present theoretical investigation \citep{Schaper:2012eb} and could explain the rapid evolution of resistant phenotypes, especially in light of the 'super-star' topology of the flow network, which has been shown analytically to magnify selection \citep{Lieberman:2005fk}. In the future we plan to investigate these open questions with an extended model, that explicitly takes into account the heterogeneity of both the CTCs and the variation in selective pressures among different organs.


\section*{Conclusion}

While the Primary and Secondary seeding hypotheses presented in this paper have been stated before, the relative probabilities between the two and the importance of the differences have yet to be presented or experimentally elucidated. We present a mathematical investigation of these differences and the theoretical implications of those differences, and have shown that primary tumors whose shed CTCs first encounter the lung are much more likely to benefit from the process of self-seeding. However, to truly understand this process, the results of specifically designed experiments need to be brought to light.  These experiments are not beyond the current technology, but have not been performed because of the lack of a theoretical construct by which to understand their results.  Our investigation and predictions, coupled with the conceptual framework presented by Scott et al. in which the vascular system was first represented as a network in the setting of metastatic spread \citep{scottfilterflow}, provide the rationale for several simple experiments which would definitively answer these questions.  

To this end, there are a number of simple to perform experiments which could shed light on not only the estimated parameters in this study but also on the process itself.  Most published literature correlating outcomes with CTC load utilize systemic venous blood, and we assert that these measurements are far too late in the process.   Early measurement and characterization of arterial circulating tumour cell populations (cells which have not yet been removed by filtration during passage through the lung capillary bed), which are easily accessible from most patients at the time of resection of their primary, could be correlated with outcome and subsequent metastatic populations.  This information could yield invaluable information about metastatic potential and which cell populations within the primary are of concern (and are targetable).   Further, simple comparison of circulating tumour cell numbers in the arterial system compared with venous system in patients with known and unknown metastatic status could extend our understanding of the biology of filtration - helping us to understand which tumors do and do not follow physically solvable metastatic patterns.  Further, emerging techniques from other disciplines, such as the evolutionary inference from phylogenetic trees as presented by \cite{Gerlinger:2012fs}, could reveal the first appearance of clones in metastatic colonies that subsequently 'fed' the primary tumor.  These and many other questions remained unanswered and the first step is creation of a robust theoretical construct on which to base further enquiry - the first steps towards which we have presented here.

\section*{Acknowledments}
This work was partially supported by the NIH/NCI Integrative Cancer Biology Program (U54 CA113007) (ARAA, PG and DB) and the NIH Loan Repayment Grant (JGS). The work of PG was partially funded by Stiftelsen Assar Gabrielssons Fond and Cancerfonden.

\bibliography{selfseed.bib}
\bibliographystyle{natbib}

\end{document}